\def\vinf{{v_{\infty}}}

\def\msun{{\rm\,M_\odot}}

\documentclass[traditabstract]{aa}  

\usepackage{graphicx}
\usepackage{txfonts}
\usepackage{lipsum}
\usepackage{lscape}                                           
\usepackage{placeins}           

\begin{document}

   \title{Capture of interstellar objects during stellar encounters}

   \author{Sean~N. Raymond\inst{1},
   Nathan~A. Kaib\inst{2}
        \and Matthew~J. Hopkins\inst{3}  }

   \institute{Laboratoire d'Astrophysique de Bordeaux, CNRS and Universit{\'e} de Bordeaux, All{\'e}e Geoffroy St. Hilaire, 33615 Pessac, France
            \email{rayray.sean@gmail.com}
            \and Planetary Science Institute, 1700 E. Fort Lowell, Suite 106, Tucson, AZ 85719, USA
                \and Te Kura Mat{\=u}, University of Canterbury, Private Bag 4800, Christchurch 8140, New Zealand
                }

\date{Submitted April 29, 2026 -- Accepted July 10, 2026}

 
  \abstract{
As they orbit within the Galaxy, stars swim through a vast population of interstellar objects (ISOs). In this paper, we use N-body simulations to show that a fraction of ISOs within $\sim$1 pc of the Sun (its tidal radius) may be captured during the flyby of another star -- a mechanism that requires no planets. Capture is most efficient when the impulse imparted by the flyby is comparable to the escape speed at the widest stable orbit, which is roughly 0.1 km/s for the Sun. ISO capture is dominated by the few highest-impulse stellar flybys,  typically involving relatively slow encounters with massive stars. Most ISOs are captured in the outer parts of the Oort cloud, with semimajor axes greater than $\sim$50,000 au. Using Monte Carlo simulations, we show that the Sun underwent only a small number of ISO-capturing flybys in its history (median [mean] of 1 [1.7]). Using the {\=O}tautahi-Oxford population model, we estimate that a few times $\sim$$10^{4}$ `Oumuamua-sized ISOs were likely captured by the Sun. This only represents a $\sim$$10^{-8\pm1}$ contribution to the total Oort cloud population, yet it contains roughly as many present-day captured ISOs as Jupiter-assisted capture provides. Given that flybys are unavoidable in the Galactic field, most stars should host sparse Oort clouds populated with ISOs captured during stellar flybys. Massive stars are both the main drivers of capture when they fly by a given star, and more efficient at capturing ISOs around themselves than low-mass stars. }
   \keywords{interstellar objects -- Oort cloud -- comets -- stellar encounters}
   \maketitle
   \nolinenumbers

\section{Introduction}

While their existence had been hypothesized for decades~\citep[e.g.][]{whipple75,mcglynn89,kresak92,sen93,moromartin09,jura11,engelhardt17}, the discovery of interstellar objects (ISOs) still contained many surprises.  Analysis of the first ISO, the 100~m-scale 1I/`Oumuamua~\citep{meech17}, hinted that such objects were far more common than predicted by simple models of planetary formation and ejection, suggesting that they may not follow an asteroid belt-like size-frequency distribution~\citep[e.g.][]{moromartin18,raymond18c,raymond18,portegieszwart18,issi19}.  To explain its discovery, \cite{do18} inferred a number density of $n \approx 0.1$ `Oumuamua-sized objects per cubic au. As 2I/Borisov was discovered by an amateur and not in the confines of a survey~\citep[e.g.][]{jewitt19}, it is challenging to use it to estimate the underlying population density.  Given its activity, 3I/ATLAS has been observed by a large number of telescopes, and \cite{hui26} derived $n \approx 3 \times 10^{-3} \, {\rm au^{-3}}$ from its detection.  

Stars can gravitationally capture ISOs onto bound orbits. Capture requires a change in an ISO's velocity comparable to its velocity at infinity relative to the Sun -- that is, its speed after subtracting off the Sun's gravitational acceleration. The best-studied ISO capture mechanism is the result of Jupiter~\citep{valtonen82,hands20,napier21,dehnen22b}. Jupiter's gravitational influence can temporarily capture passing ISOs, whose orbits remain stably within the confines of the Solar System for millions of years (or more), before being re-ejected~\citep{napier21b,dehnen22}.  Other mechanisms for ISO capture have been studied to some degree -- for instance, the Galactic tide has been shown to temporarily capture passing ISOs~\citep{penarrubia23}. 

In this paper, we show that stellar flybys are another mechanism for ISO capture -- and, unlike other mechanisms, one that does not require the existence of any giant planets.  During a stellar encounter, the impulse imparted by the flyby star can lead to the capture of a fraction of ISOs that happen to be passing through a star's sphere of gravitational influence (defined by the tidal radius -- see Eq. 1), and that also have a sufficiently low velocity relative to the Sun. This mechanism is analogous to the capture of the irregular satellites of the giant planets during close encounters suffered during the giant planet instability~\citep{nesvorny07,nesvorny14}.  

We first present a series of N-body simulations of stellar flybys including $10^7$ ISO particles each, and use them to devise a simple criterion for capture.  We determine the orbital characteristics of ISOs and calibrate to the {\=O}tautahi-Oxford ISO population model~\citep{hopkins23,hopkins25,dorsey25}. Then,  we perform a series of Monte Carlo realizations of stellar flybys of the Sun within the Galactic field, to estimate the number of ISOs captured throughout its lifetime. We discuss how our results can be extrapolated to other stars, as well as the limitations of our study.

\section{N-body simulations of ISO capture}

Our simulations start with a population of ISOs passing through the Sun's sphere of influence, and simulated a single stellar flyby with predefined characteristics.  Each simulation included a population of $10^7$ ISOs with initially isotropic velocity and spatial distributions, and a velocity at infinity $\vinf$ distribution evenly spread between zero and 100 km/s.\footnote{In practice, we ran $10^4$ simulations with $10^3$ ISOs each and an identical flyby (in `embarrassingly parallel' fashion).  Given that ISOs do not self-interact, we combined the results. } This allows us to determine the capture rate as a function of initial velocity and to convolve it with a more realistic distribution (see below).  

We used 1 pc (206265 au) as the outer bound of our simulations.  This is only slightly larger than estimates of the Sun's tidal radius $R_{tide}$, beyond which an object is no longer bound to its star~\citep{smoluchowski84}, defined as~\citep{tremaine93}:
\begin{equation}
    R_{tide} \approx 1.9 \times 10^5 \mathrm{au} \, \left(\frac{M_\star}{\msun}\right)^{1/3} \rho_0^{-1/3},
\end{equation}
\noindent where $M_\star$ is the stellar mass and $\rho_0$ is the Galactic density relative to the local value of roughly $0.1 \msun \ {\rm pc}^{-3}$~\citep{holmberg00,chakrabarti21}.  When a given ISO passed beyond 1 pc, it was re-initialized at a different point on a sphere of 1 pc in radius, with the same $\vinf$, with a trajectory chosen to scale with the square of the impact parameter~\citep{henon72}.  

We then subjected the Sun-ISO system to a single stellar flyby, with varying parameters.  The star was initialized just interior to the 1 pc sphere, and integrated as it passed by the Sun.  Our simulations were evolved with the {\tt Mercury} integrator~\citep{chambers99}, modified to account for the displacement of ISOs that leave our 1 pc sphere, as described above.  Each system was integrated with the Bulirsch-Stoer integrator~\citep{press92} with accuracy parameter of $10^{-15}$ and a default timestep of 100 days for long enough for the star to pass back outside of 1 pc (20-100 kyr depending on $\vinf$).  

\begin{figure}
	\includegraphics[width=0.9\columnwidth]{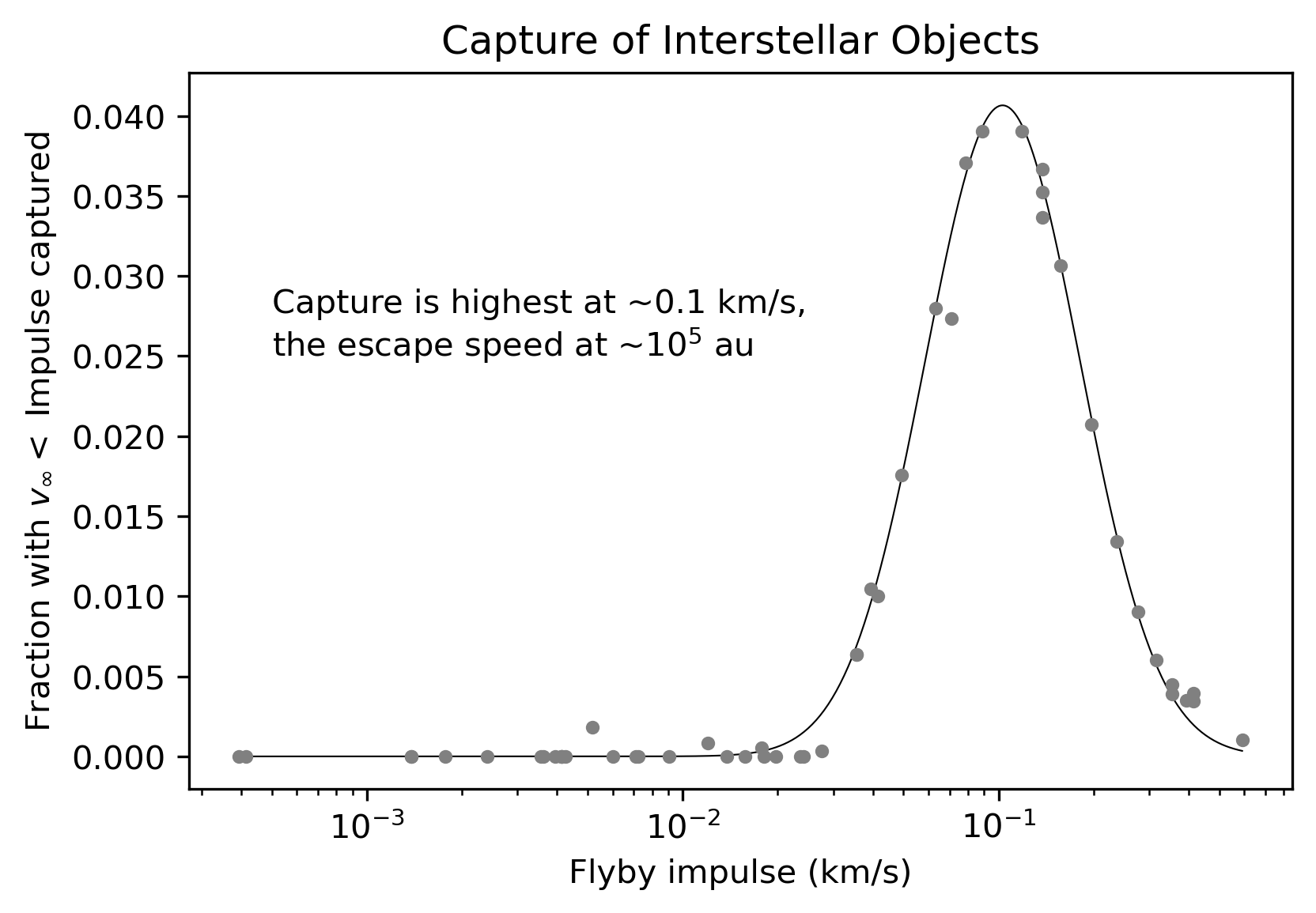}
 \caption{The fraction of ISOs with $\vinf$ less than the flyby impulse that are captured in a given flyby, as a function of the impulse. Each symbol is a simulation with $10^7$ ISOs.  The grey curve is a lognormal fit to the simulations.
 }
    \label{fig:capture_prob}
\end{figure}

There are three relevant flyby parameters: the stellar mass $M_\star$, the impact parameter $b$, and the velocity at infinity $\vinf$. In a preliminary set of runs, we tested parameter space more or less at random.  Then, we performed a systematic sweep of flybys that appeared capable of capturing ISOs.  We fixed a fiducial case with $M_\star = 0.7 \msun$, $\vinf = 30$~km/s, and $b = 3000$~au.  We then did a parameter sweep, fixing $M_\star = 0.7 \msun$ and $\vinf = 30$~km/s, then testing $b =$ 100, 300, 1000, 10000, 30000, and $10^5$~au.  Next, we fixed $M_\star = 0.7 \msun$ and $b = 3000$~au, and tested $\vinf =$1, 3, 10, and 100 km/s.  Finally, we fixed $\vinf = 30$~km/s and $b = 3000$~au and tested $M_\star =$ 0.02, 0.07, 0.2, 0.8, 1, 1.4, 2, 2.5, 3.2, 4, 6, 7, 8, 10, 12, 14, 16, 18, 20, and $30 \msun$. We consider an ISO to be captured if its aphelion distance is less than our adopted tidal radius of 1 pc.  

The number and distribution of captured ISOs scales with a single quantity: the flyby impulse, defined as:
\begin{equation}
\rm{Impulse} = \frac{2 G M_\star}{\vinf b}, 
\end{equation}
\noindent where $G$ is the gravitational constant. The ISOs that were captured in our simulations had $\vinf$ smaller than the flyby impulse. This demonstrates that the capture mechanism was indirect: the flyby star's velocity impulse to the Sun caused a fraction of ISOs within the Sun's Hill sphere to become bound.  Of course, only ISOs with the appropriate velocity vectors were susceptible to capture. 

Figure~\ref{fig:capture_prob} shows the capture efficiency of ISOs that were captured as a function of the flyby impulse. For impulses lower than 0.025-0.03 km/s, virtually no ISOs are captured.  There is a steady increase in the capture probability until it reaches a peak at an impulse of $\sim 0.1$~km/s, then decreases at higher impulse. The tightness of the relation in Fig.~\ref{fig:capture_prob} -- which combines simulations spanning three orders of magnitude in stellar mass, impact parameter, and encounter velocity -- demonstrates that the impulse is the controlling parameter irrespective of stellar mass, encounter geometry, or velocity.

The Solar System's tidal radius sets the natural velocity scale for capture. An object that is securely bound, with eccentricity below unity, will typically have a semimajor axis of order $a \approx 10^5$ au, corresponding to an escape speed of $\sim$0.1 km/s. Capture therefore requires a change in velocity of this same order, comparable to that needed for ejection. At lower impulses, the perturbation is insufficient to bind the object, although some ISOs can be tenuously captured for impulses down to a fraction of this value. Capture is most efficient when the impulse is comparable to the escape speed at the tidal radius. At higher impulses, objects can in principle be captured onto tighter orbits, but the effective capture cross section decreases: stronger encounters require more specific geometries and therefore occur less frequently. As a result, the overall capture efficiency declines at large impulses. In this way, the capture probability naturally peaks at an impulse comparable to the escape speed.

The capture probability from Fig.~\ref{fig:capture_prob} can be fit with a lognormal distribution:
\begin{equation}
P(x) = A \, H(x - x_{\min}) \,
\exp\left[
- \frac{\left(\ln x - \ln x_0\right)^2}{2\sigma^2}
\right],
\end{equation}
\noindent where $x$ is the flyby impulse, $H$ is the Heaviside function (equal to zero below $x_{\min}$ and 1 above), $x_0$ is the characteristic impulse at which capture is maximized, and $\sigma$ describes the width of the capture window in log-space.  In the fit shown in Fig.~\ref{fig:capture_prob}, $A$ = 0.040, $x_0$ = 0.105 km/s, $\sigma$ = 0.55, and $x_{\min}$ = 0.02 km/s.

\begin{figure}
	\includegraphics[width=0.9\columnwidth]{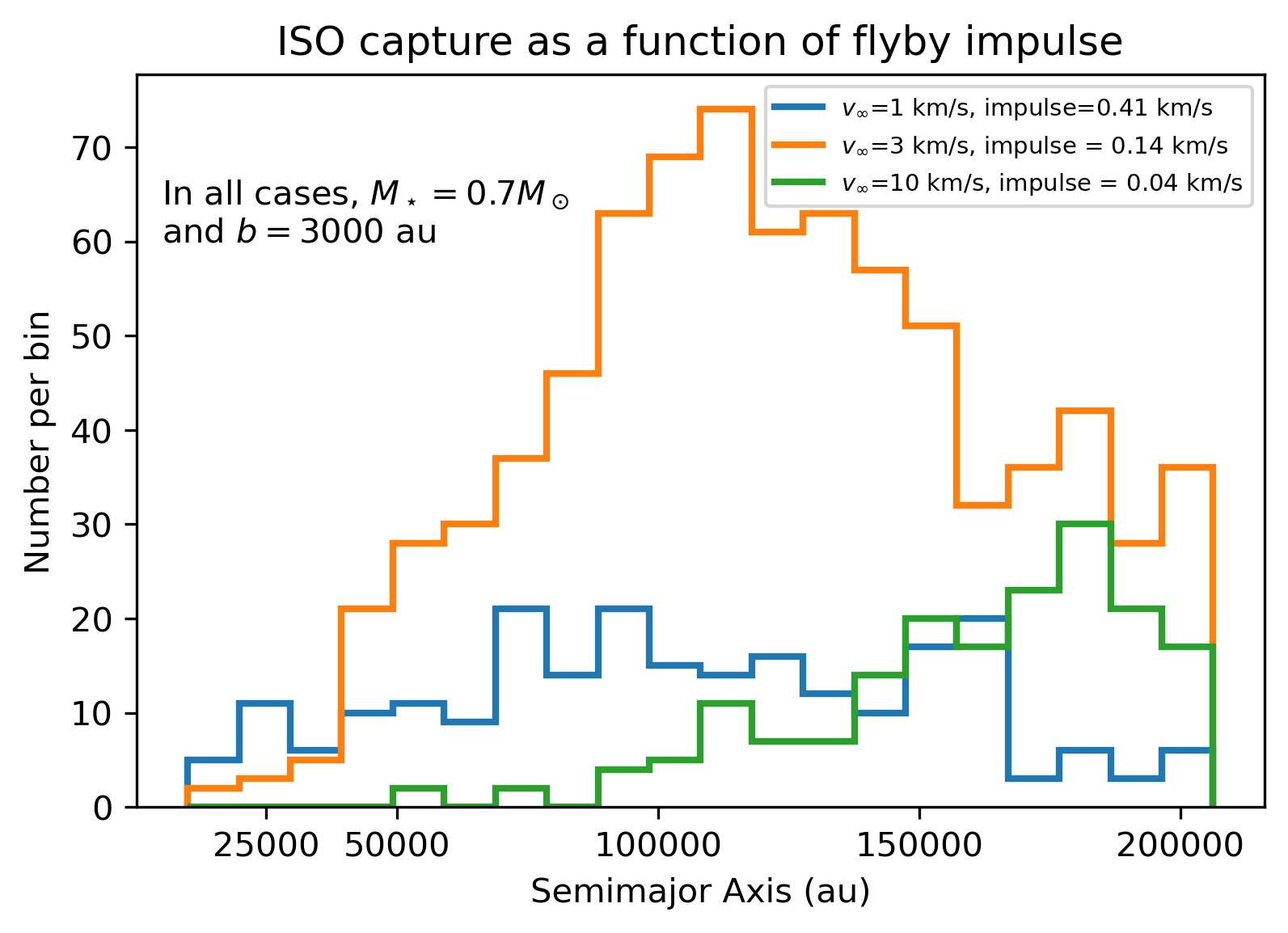}
 \caption{Semimajor axis distribution of captured ISOs in three simulations, each with stellar mass $M_\star = 0.7 \msun$ and impact parameter $b = 3000$~au, but with velocity at infinity $\vinf =$~1, 3, and 10 km/s. }
    \label{fig:capture_vinf}
\end{figure}

Figure~\ref{fig:capture_vinf} illustrates the impulse dependence of ISO capture using three simulations, each with $M_\star = 0.7 \msun$ and $b = 3000$~au, but with $\vinf =$~1, 3, and 10 km/s. Recall from Eq. 2 that the flyby impulse scales inversely with $\vinf$. The simulation with $\vinf =3$~km/s had an impulse of 0.14 km/s, at the peak from Fig.~\ref{fig:capture_prob}. ISOs captured in that simulation peaked in semimajor axis just above $10^5$.  The lower-impulse flyby in Fig.~\ref{fig:capture_vinf} (with $\vinf = 10$~km/s) captured fewer total objects and these were skewed to larger semimajor axis, where a smaller impulse is needed. The higher-impulse flyby also captured fewer total ISOs, but they were skewed to lower semimajor axes.

The question becomes, how many ISO-capturing flybys has the Sun undergone since its formation?  To address this, we must model the Sun's encounter history within the Galaxy.

\section{Monte Carlo simulations of flybys and ISO capture}

We built a simple Monte Carlo model of stellar flybys, following the approach of \cite{rickman08}.  The masses of flyby stars were chosen to follow the observed mass function of nearby stars determined by \cite{reid02}, with a lower limit set to $0.05 \msun$.  The orbital velocities of stars were chosen as a function of their masses to match the present-day velocity distribution~\citep[see Table 1 of ][]{rickman08}.  The resulting velocity distribution has a median of 40.2 km/s~\citep[consistent with][]{binney08} and a standard deviation of 16.9 km/s when averaged over all stellar types. We use these distributions to calculate a flux-weighted encounter rate, which is itself a function of stellar mass.  We find that, over all stellar types, the Sun undergoes roughly one flyby within 1 pc every 50,000 years. This rate is consistent with previous studies~\citep[e.g.][]{heisler86,zink20,brown22} and implies that the Sun has undergone roughly 90,000 stellar flybys that penetrated its tidal radius during its 4.5 Gyr lifetime.  Armed with these quantities, our code generates the appropriate number of encounters with realistic mass, velocity, and close approach distributions.

\begin{figure}
	\includegraphics[width=0.9\columnwidth]{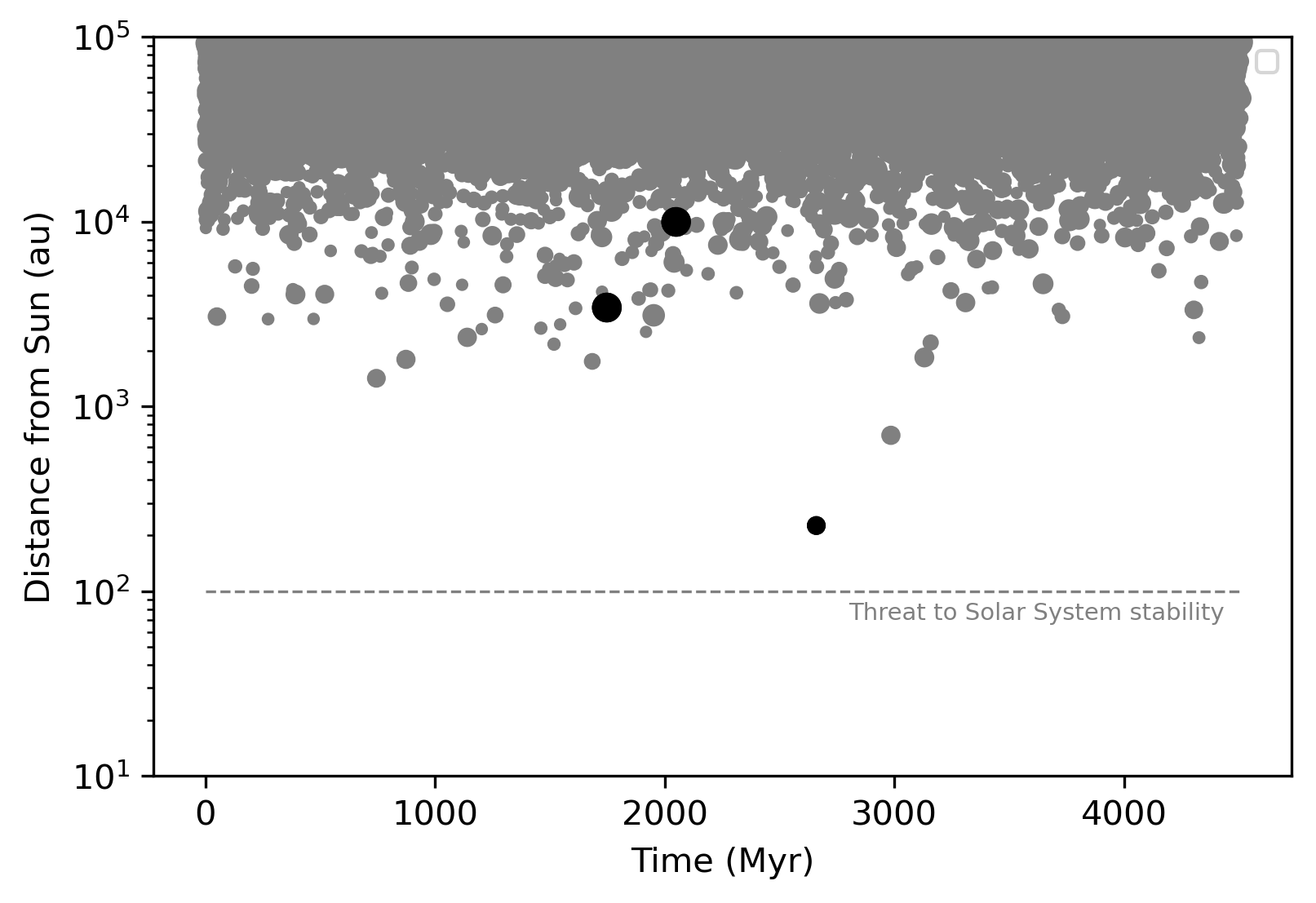}
 \caption{An example encounter history of the Sun. Each symbol shows a given stellar flyby, with a size that scales with the stellar mass.  The black symbols imparted an impulse sufficient to induce ISO capture.  
 }
    \label{fig:MC_example}
\end{figure}

Figure~\ref{fig:MC_example} shows an example flyby history of the Sun drawn with this Monte Carlo method. In this case, there were three encounters strong enough to capture ISOs (shown in black); these were not necessarily the closest encounters, as a slow-moving, high-mass star can impart a strong impulse at a large distance (see Eq. 2). Nonetheless, the highest-impulse flybys tend to come from stars significantly more massive than the Sun~\citep[][]{rickman08}.  In this example, the first two ISO-capturing flybys were each with a star of $4 \msun$, and, coincidentally, both had flybys with $\vinf \approx 27$~km/s. The first flyby had a much closer approach to the Sun (impact parameter $b = 3432$ au for the first vs. $b = 9972$~au for the second), and imparted an impulse of 0.075 km/s compared with 0.026 km/s for the second flyby.  The third flyby was in a different realm of parameter space, consisting of a $0.47 \msun$ M dwarf passing within 226 au of the Sun with $\vinf = 72.4$~km/s. It imparted an impulse of 0.05 km/s.

To estimate the actual number of ISOs that should be captured in a given flyby, we calibrate to the {\=O}tautahi-Oxford ISO population model~\citep{hopkins23,hopkins25,dorsey25}.  At the small heliocentric speeds at which ISOs are susceptible to capture ($\vinf \lesssim 0.1$~km/s), the ISO velocity distribution probability density function is approximately uniform with a value of $3 \times 10^{-6}$~(km/s)$^{-3}$. Thus, the fraction of ISOs with heliocentric velocity at infinity less than $\vinf$ is:
\begin{equation}
F(\vinf)=(3\times 10^{-6}\,(\mathrm{km/s})^{-3})\cdot \frac{3\pi}{4}\vinf^3
\end{equation}
\noindent and the total number within our spherical volume $V$ is $F(\vinf ) n V$, where $n$ is the (poorly-constrained) number density of ISOs in the Galactic neighborhood. \cite{do18} calculated $n \approx 0.1\, {\rm au^{-3}}$ from the detection of 1I and \cite{hui26} derived $n \approx 3 \times 10^{-3} \, {\rm au^{-3}}$ from the detection of 3I.  These two objects are more than an order of magnitude different in size, and also vary greatly in the parameter regimes in which they are detectable, given that one is active and the other is not~\citep{engelhardt17}. 

The three strongest-impulse flybys from Fig.~\ref{fig:MC_example} captured a total of 431,000 ISOs, assuming $n = 0.1\, {\rm au^{-3}}$ (appropriate for `Oumuamua-sized objects).  Yet capture varied hugely between the flybys, given the strong dependence of capture on the impulse (Fig.~\ref{fig:capture_prob}). The first flyby had the highest impulse and captured 366,000 ISOs, or roughly 85\% of the total number.  The second flyby only captured 1000 ISOs and the third 64,000.  If we instead adopt $n = 3 \times 10^{-3} \, {\rm au^{-3}}$, these flybys captured a total of $\sim$~13,000 3I/ATLAS-sized objects. 

We performed a set of $10^3$ Monte Carlo realizations to evaluate the likelihood of different outcomes. It's worth noting that some of these realizations are incompatible with the present-day Solar System.  Flybys within $\sim$100~au put the dynamical stability of the planets at risk~\citep{brown22,kaib25} -- 21 of our Monte Carlo realizations had such a close encounter, and we exclude those from further analysis.  The median closest flyby suffered by the Sun was $\sim$630 au. The median system underwent one flyby with impulse large enough to capture ISOs, with a mean of 1.66 ISO-capturing flybys and a range of zero to 7, with 18\% of realizations having no ISO-capturing flyby. The median maximum impulse was 0.04 km/s, in a range consistent with previous work by \cite{kaib09}.  This value for the impulse is located along the increasing slope in capture probability seen in Fig.~\ref{fig:capture_prob}. An additional 18\% of realizations suffered a flyby with an impulse at or above the peak in capture probability at $\sim$0.1 km/s, without passing so close as to disrupt the planets' orbits. The median mass of all ISO-capturing flybys was $1.22 \msun$, with a median impact parameter of 2300~au and a median $\vinf$ of 28 km/s.  The median number of captured `Oumuamua-sized ISOs (assuming $n = 0.1\, {\rm au^{-3}}$) was $2.1 \times 10^4$, with a range from zero to $1.1 \times 10^7$ and a 20th to 80th percentile range of $10^3$ to $3.3 \times 10^5$. 

Massive stars play an outsized role in the capture of ISOs.  Of the thousands of ISO-capturing flybys in our Monte Carlo realizations, 59\% involved the flyby of a star more massive than the Sun.  Only 13\% were the result of a flyby with $M_\star < 0.5 \msun$, and only 3\% with $M_\star < 0.25 \msun$.  While low-mass stars dominate by number, they play only a minor role in accelerating stars and causing ISO capture.

\section{Captured ISOs within the Oort cloud}

\begin{figure}
	\includegraphics[width=0.9\columnwidth]{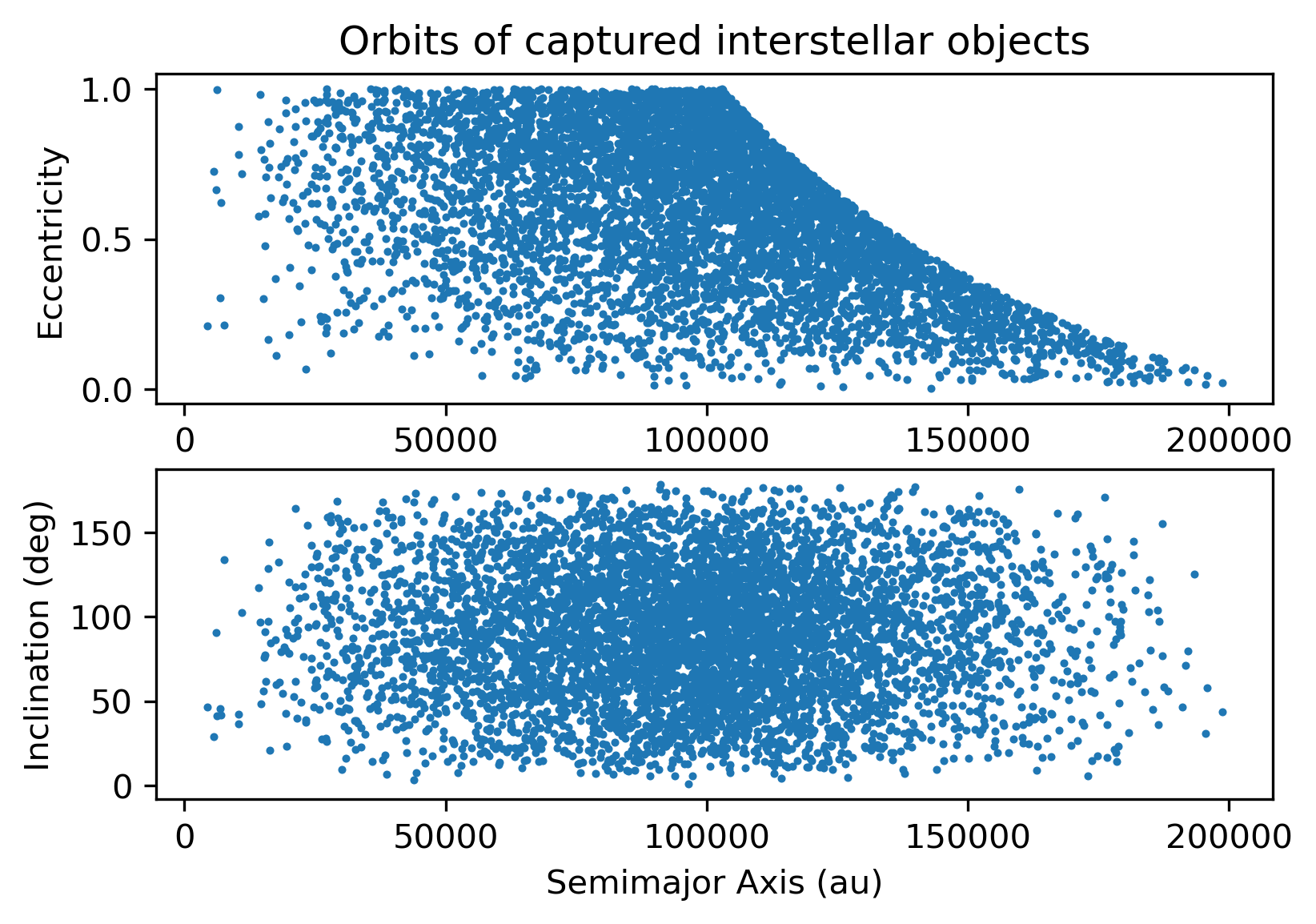}
 \caption{Orbital distribution of captured ISOs across all of our simulations. }
    \label{fig:capture_orbits}
\end{figure}

Figure~\ref{fig:capture_orbits} shows the orbits of all ISO particles captured in our different simulations. There is clear edge in eccentricity coming from the imposed requirement that a captured ISO must have an aphelion smaller than 1 pc. The overall distribution is clearly reminiscent of that of the Oort cloud, as captured ISOs have an isotropic inclination distribution and a roughly thermal eccentricity distribution (with a median eccentricity of $\sim 0.72$ for semimajor axes below $10^5$~au -- to avoid the bias from the aphelion requirement). 

Yet the number of captured ISOs is tiny compared with native Oort cloud comets. Our simulations imply that the Sun may have captured a few times $10^4$ 100 m-scale ISOs, or $\sim 10^3$ km-scale ISOs. This is 7-9 orders of magnitude less than the commonly-expected Oort cloud population~\citep[see review in][]{kaibvolk24}.  \cite{boe19} estimated that there are on the order of $10^9$ long-period comets larger than 1 km in the Solar System.  Thus, even if all of the captured ISOs were on inner Solar System-crossing orbits, they would still be outnumbered by native comets at a rate of $10^6$ to one. The catalog of long-period comets needs to approach that order to have a reasonable hope of finding an ISO captured by the mechanism described here.

The most favorable case would be if there was a rapid influx of such objects shortly after their capture, rather than a steady state.  However, a flyby would also create a spike in native cometary populations~\citep{hills81,hut87}, so it's not clear that this would actually help.

Even though captured ISOs will clearly be vastly outnumbered by native comets, it is possible that they could be distinguishable by their chemical and isotopic compositions.  Indeed, both 2I and 3I have compositions that are significantly different from Solar System comets~\citep[e.g.][]{bodewits20,cordiner26,opitom26}.  There are a handful of long-period comets with unusual compositions~\citep{fink92,schleicher08,mckay19}.  However, our results suggest that it is statistically very unlikely that these objects are ISOs captured in stellar flybys. 

The orbits onto which ISOs are captured -- mostly beyond $5 \times 10^4$au (Fig.~\ref{fig:capture_orbits}) -- are also the most fragile in the Solar System. Objects at these distances are strongly perturbed by the Galactic tide and by subsequent stellar flybys, and native outer Oort cloud bodies on comparable orbits are eroded on timescales of a few Gyr~\citep[e.g.][]{kaib11b,torres19}. Captured ISOs should be lost at a similar rate. However, because ISO-capturing flybys are distributed roughly uniformly in time, objects captured in the recent past have had little time to be removed. At any epoch, the bound captured population therefore reflects a balance between episodic capture and continuous erosion; the numbers quoted above represent objects ever captured, of which we assume roughly half survive to the present day (see Sec. 5.3).

\section{Discussion}

\subsection{Extrapolation to other stars}

Flyby-driven capture of interstellar objects is a generic mechanism that must apply to all stars, whether or not they have planets of their own. The dynamics is simple to extrapolate -- all Solar-mass stars in the Solar neighborhood should have a similar population of ISOs captured by this process as the Sun.  

There is a stellar mass dependence to ISO capture.  Eq. 1 shows that the tidal radius $R_{tide} \propto M_\star^{1/3}$.   The critical impulse to capture an object -- essentially, the velocity that should correspond to the peak in capture efficiency from Fig.~\ref{fig:capture_prob} -- scales with the escape speed at the tidal radius.  This means that the critical impulse velocity, $v_{imp,crit} \propto M_\star^{1/3}$. The number of ISOs available to capture, meanwhile, scale with the volume enclosed by the tidal radius, such that $N_{iso} \propto R_{tide}^3 \propto M_\star$.  Considering just these two factors, the capture rate of ISOs should scale strongly with stellar mass, as $M_\star^{4/3}$. This implies that higher-mass stars are likely to have a substantially richer population of captured ISOs than lower-mass ones.  Nonetheless, we expect that most stars should, at any given time, host a sparse population of ISOs captured from stellar flybys into their Oort clouds.

\begin{table*}
\caption{Comparison of the two ISO capture channels.}
\label{tab:capture_channels}
\centering
\begin{tabular}{l l l}
\hline\hline
 & Jupiter-assisted capture & Stellar flyby capture (this work) \\
\hline
Requirement & Giant planet(s) & None -- operates for any star \\
Capture tempo & Quasi-steady, $\sim$2 per 1000 yr\tablefootmark{a} & Episodic; dominated by $\sim$1--2 strong flybys per lifetime \\
Total captured over 4.5 Gyr\tablefootmark{a} & $\sim 9 \times 10^{6}$ & $\sim 2 \times 10^{4}$ (median; large stochastic spread) \\
Typical captured orbits & Loosely bound, perihelia near Jupiter & Outer Oort cloud, $a \gtrsim 5\times10^{4}$ au \\
Residence time & $\sim$1--10 Myr\tablefootmark{a} & Up to Gyr (limited by Galactic tide and later flybys) \\
Present-day captured population & $\sim 10^{3}$--$10^{4}$ & $\sim 10^{4}$ (assuming $\sim$50\% retention) \\
\hline
\end{tabular}
\tablefoot{
Numbers assume `Oumuamua-sized objects~\citep[$n \approx 0.1$~au$^{-3}$;][]{do18}.
\tablefoottext{a}{\citep{dehnen22}.}
}
\end{table*}

\subsection{A candidate ISO-capturing flyby}

The Sun underwent the relatively close flyby of the Sun-like star HD 7977 roughly 2.8 Myr ago~\citep{bailerjones22}.  The closest approach is hard to constrain using Gaia, but there is a $\sim$5\% chance it passed within 4000 au.  Using the dynamics and alignment of different categories of comets, \cite{kaib26} estimated its closest approach at 6000-10000 au.  Given that the star's mass and $\vinf$ are relatively well measured (at $1.1 \msun$ and 27 km/s, respectively), we can establish that this was probably not a strong ISO-capturing event.  The maximum plausible impulse -- if the star had a closest approach of 3000 au from the Sun -- would have been 0.024 km/s, just below the threshold for ISO capture.  We performed a series of N-body simulations which showed, as expected, that ISO capture was unlikely during this event.  

\subsection{Comparison with other ISO capture mechanisms}

Jupiter-assisted ISO capture has been demonstrated as a viable mechanism~\citep{valtonen82,hands20,napier21,dehnen22b}.  \cite{dehnen22} showed that this acts to capture roughly 2 new `Oumuamua-sized ISOs per 1000 yrs, which amounts to $9 \times 10^6$ captures over the Sun's lifetime.  This is a factor of $\sim$500 higher than the time-averaged capture rate from stellar flybys, if we assume the median number of captured objects from the Monte Carlo simulations presented in Section 3. Yet ISOs captured through stellar flybys are likely to be much longer-lived than those captured with Jupiter's assistance.  \cite{dehnen22} estimated the residence time of captured ISOs in the $\sim 1-10$~Myr range, as scattering by Jupiter and Saturn rapidly leads to their ejection.  If we make the simple assumption that half of flyby-captured ISOs survive to the present-day~\citep[e.g.][]{kaib11b}, then the number of ISOs residing in the Solar System that were captured by each mechanism should be similar, albeit on quite different orbits. 
\cite{penarrubia23} showed that the Galactic tide creates a halo of almost-captured ISOs around the Solar System.  While this halo is relatively diffuse, containing $\sim 10^7$ objects, one can imagine that this population of very low relative-velocity objects could significantly increase the capture efficiency of stellar flybys.  \cite{forbes26} also found a small (factor $\sim$1/100) excess of low-velocity ISOs, coming in large part from objects that were barely ejected from the Solar System. 

\subsection{Limitations of our work and Uncertainty budget}

The number of captured ISOs is uncertain for several reasons, in decreasing order of importance. The first is the stochastic nature of the Sun's encounter history: across our Monte Carlo realizations, the number of captured `Oumuamua-sized ISOs from zero (18\% of realizations) to $7.1\times10^6$, with a median of $1.8\times 10^4$ and a 20th-80th percentile range of $10^3$ to $3.3 \times 10^5$. Because capture is dominated by the single strongest flyby, this distribution is heavy-tailed. The second is the ISO number density, which is uncertain at the order-of-magnitude level (Do et al. 2018; Hui et al. 2026) and enters our results linearly. The third is the shape of the low-velocity tail of the ISO velocity distribution (see Sect. 5.4). In calculating the absolute number of captured ISOs, we have relied exclusively on the {\=O}tautahi-Oxford ISO population model~\citep{hopkins23,hopkins25,dorsey25}.  Even a modest change in the velocity distribution of these objects, especially at very low velocities, would have a considerable impact on our results.  For instance, if the abundance of ISOs at very low relative velocity were to be substantially higher~\citep[perhaps as a result of the Galactic tide;][]{penarrubia23,forbes26}, then the number of captured ISOs could be substantially larger. Finally, our adopted capture criterion and finite simulation volume introduce uncertainties at the factor-of-order-unity level.

Finally, our simulations are admittedly simplistic, as we ignore the Sun's orbit within the Galaxy.  Allowing the Sun's orbital radius to oscillate would make $R_{tide}$ vary as a function of time, which would likely lead to loss of the outermost orbits of the Oort cloud~\citep[e.g.][]{kaib11b}.  Our simulation bounding box is likewise smaller than one would like; however, there is an $R^3$ cost to to increasing it, in terms of the number of ISOs to simulate or the number of simulations to run.  

\subsection{Universality of flyby-driven capture}
The Solar System has long been suspected of containing captured, non-native material. Stellar encounters have been invoked to capture planetesimals from a passing star onto Sedna-like orbits~\citep[e.g.][]{kenyon04,jilkova15}, and it has been proposed that a substantial fraction of the Oort cloud itself may have been captured from the Sun's siblings in its birth cluster~\citep{levison10}. The recent ISO discoveries have now confirmed directly that extraterrestrial material regularly passes through the Solar System. The mechanism presented here adds a capture pathway that requires no birth cluster and no giant planets, operating in the Galactic field throughout a star's lifetime.  While this mechanism is only activated during sufficiently strong stellar flybys, of which there have likely only been 1-2 during the Sun's history, it is likely responsible for a large fraction of the captured ISOs currently residing in the Solar System (see Table 1). 

\section{Summary}

To summarize, this paper contains three main results. First, ISO capture during stellar flybys is controlled by a single parameter -- the flyby impulse (Eq. 2) -- and is most efficient when the impulse is comparable to the escape speed at the tidal radius, roughly 0.1 km/s for the Sun (see Fig.~\ref{fig:capture_prob}). Second, capture is dominated by the few strongest flybys in a star's history, which preferentially involve massive stars; massive stars are also more efficient at capturing ISOs around themselves. Third, the Sun probably underwent only 1-2 capture-inducing flybys, and has likely captured a few times $10^4$ `Oumuamua-sized ISOs, predominantly onto orbits in the outer Oort cloud -- a contribution of only $\sim 10^{-8}$ of the total Oort cloud population, but one that essentially every field star should share.

\begin{acknowledgements}
We thank the anonymous referee for a constructive report that improved the paper.
NAK's contributions to this work were supported from NASA Solar System Workings grant 80NSSC24K1874 and NASA Emerging Worlds grant 80NSSC23K0771.
MJH appreciates support from the Elaine P. Snowden Fellowship.
SNR acknowledges funding from the Programme Nationale de Planetologie (PNP) of the INSU (CNRS), and in the framework of the Investments for the Future programme IdEx, Universite de Bordeaux/RRI ORIGINS.
\end{acknowledgements}


\end{document}